# Two Distinct Modes of Hydrogen-Bonding Interaction in the Prototypic Hybrid Halide Perovskite, Tetragonal $CH_3NH_3PbI_3$

**June Ho Lee,**[†] **Jung-Hoon Lee,**[†] **Eui-Hyun Kong,**[‡] **and Hyun M. Jang\***[†]

[†] Department of Materials Science and Engineering, and Division of Advanced Materials Science (AMS), Pohang University of Science and Technology (POSTECH), Pohang 790-784, Republic of Korea

[‡] Korea Atomic Energy Research Institute (KAERI), Yuseong-Gu, Daejeon 305-353, Republic of Korea

**Abstract:** In spite of the key role of hydrogen bonding in the structural stabilization of the prototypic hybrid halide perovskite, $CH_3NH_3PbI_3$ ($MAPbI_3$), little progress has been made in our in-depth understanding of the hydrogen-bonding interaction between the $MA^+$-ion and the iodide ions in the $PbI_6$-octahedron network. Herein, we show that there exist two distinct types of the hydrogen-bonding interaction, naming $\alpha$- and $\beta$-modes, in the tetragonal $MAPbI_3$ on the basis of symmetry argument and density-functional theory calculations. The computed Kohn-Sham (K-S) energy difference between these two interaction modes is 45.14 meV per MA-site with the $\alpha$-interaction mode being responsible for the stable hydrogen-bonding network. We have further estimated the individual bonding strength for the ten relevant hydrogen bonds having a bond critical point. The net difference in the total hydrogen-bonding energies between these two interaction modes is 43.87 meV per MA-site, which nearly coincides with the K-S energy difference of 45.14 meV.

# 1. INTRODUCTION

Organic-inorganic hybrid perovskite-based solar cells have revolutionized the photovoltaic landscape[1-9] as they have demonstrated unprecedentedly high power conversion efficiencies (PCEs) and low cost. Their electrical PCE increases extremely rapidly and has reached ~19 % in 2013, up from ~3 % in 2009.[1-9] The observed unusually high PCEs are currently attributed to several relevant physical factors that include low optical bandgaps,[10] large absorption coefficients,[1] and long carrier diffusion lengths.[11,12] In addition to high PCEs, hybrid halide perovskites of $RMX_3$-type show a remarkable capability of demonstrating diverse photovoltaic properties by suitable substitution or modification of organic molecules (R)[13] or metal (M) ions[14] in the hybrid perovskite lattice.

Among numerous hybrid perovskites, a group of halides having the stoichiometry of $CH_3NH_3PbX_3$ (abbreviated as $MAPbX_3$, where X = Cl, Br, I) is the most widely studied and considered to be a typical hybrid perovskite. It is known that $MAPbX_3$ undergoes consecutive phase transitions with the following sequence**:** cubic-tetragonal-orthorhombic allotropic phases with  decreasing temperature.[15,16] In case of the low-temperature orthorhombic phase, the organic molecules (MAs) are well oriented to maximize the hydrogen-bonding interaction between the MA groups and the corner-shared $PbI_6$ octahedra.[17] Thus, the orientation of the MA group and consequently the positions of hydrogen atoms are well defined in the orthorhombic phase. In the room-temperature-stable tetragonal phase or in the high-temperature cubic phase, on the contrary, the organic MA molecules are randomly oriented with no clear orientational correlation between them.[18] The configuration of the resulting hydrogen-bonding network is thus extremely complicated, which would lead to numerous local minima in the potential-energy surfaces. Thus, in the case of tetragonal or cubic phase, it seems to be extremely difficult to find the optimum configuration which corresponds to the global minimum in the energy-configuration space of $MAPbX_3$.

According to the previous theoretical study, the organic cations of different sizes and hydrogen-bonding interactions [*e.g.*, $CH_3NH_3^+$ and $(NH_2)_2CH^+$] are capable of affecting the optical bandgaps of $RPbI_3$-based perovskites.[19] Similarly, Filip *et al.*[20] have experimentally shown that tunable optical bandgaps are achieved by controlling the degree of the $PbI_6$ octahedral tilting through the steric size of the organic cation, $MA^+$. According to these two studies,[19,20] the optical bandgap can be reduced by decreasing the degree of the octahedral tilting, which, in turn, can be achieved by adjusting the degree of the hydrogen-bonding interaction between the halides and H atoms bonded to the MA group. Several other studies[21-25] also indicate the important role of the $MA^+$-ion orientation and, thus, the hydrogen-bonding interaction in controlling the core properties of the $MAPbX_3$-based perovskite solar cells, which includes the enhanced carrier diffusion length,[21] the ferroelectric photovoltaic effect,[22] and the interplay of the MA-dipole orientation with the stability of perovskite structure.[24,25]

In spite of the key role of the MA-dipole orientation and consequent hydrogen-bonding interaction, little progress has been made in our systematic understanding of (i) the stable configuration of MA$^+$-ions in the perovskite unit cell and (ii) the nature and strength of the hydrogen bonding between the MA$^+$- ion and the halide (X) ions in the PbX$_6$-octahedron network. Herein, we show that there exist two distinct types of the hydrogen-bonding interaction in the tetragonal phase which is relevant to room-temperature performance of the prototypic MAPbI$_3$-based solar cells. On the basis of symmetry argument, we will predict the existence of two distinct chemical environments for the MA$^+$-ion in the tetragonal phase and computationally show that one of these two is responsible for the stable hydrogen-bonding interaction between the MA$^+$-ion and the surrounding PbI$_6$-octahedron cages.

## 2. COMPUTATIONAL METHODS

We have performed *ab initio* density functional theory (DFT) calculations on the basis of the generalized gradient approximation (GGA)[26] implemented with projector augmented-wave (PAW) pseudopotential[27] using the Vienna *ab initio* Simulation Package (VASP).[28,29] To assess the effect of the van der Waals (vdW) interaction on the structural relaxation, we have performed all *ab initio* calculations with the vdW interaction as implemented in VASP by Grimme.[30] Most of the DFT calculations were performed by adopting (i) a 4x4x3 Monkhorst-Pack (M-P) *k*-point mesh[31] centered at the Γ-point and (ii) a 500-eV plane-wave cutoff energy. In the band-structure calculations, however, we have initially adopted a 3x3x2 M-P *k*-point mesh to obtain a relaxed structure and subsequently used a 9x9x6 ***k***-point mesh to accurately assess the ***k***-point-dependent Kohn-Sham energy. All the structural relaxations were performed with a Gaussian broadening of 0.05 eV. The ions were relaxed until the Hellmann-Feynmann forces on them were less than 0.01 eV·Å$^{-1}$. The topological analysis of electronic density contours was performed by suitably exploiting the AIM-UC program.[32]

## 3. RESULTS AND DISCUSSION

**3.1. Two Distinct Chemical Environments for the Organic-Cation Orientation.** Quarti *et al.*[24] computationally showed that a set of polar (ferroelectric-like) structures formed by a preferred MA$^+$-ion orientation is more stable, in general, than a set of apolar (antiferroelectric-like) structures formed by an isotropic distribution of the MA dipoles, which indicates an important role of the MA$^+$-ion orientation in the stability of the perovskite lattice. Molecular dynamics computations[33] and first-principles study[34] further showed that for both cubic and tetragonal phases, the MA$^+$ cations are oriented parallel to the facial direction of the inorganic cage. On the basis of these theoretical studies, one can describe the orientation of the organic cations (MA$^+$) in the tetragonal phase by eight different initial orientations.[24] These 8 orientations are described by two characteristic angles, $\theta$ and $\phi$, and are graphically illustrated in Figure 1a, where *a*, *b*, and *c* denote the three crystallographic axes of the tetragonal perovskite lattice. In the figure, $\theta$ defines the angle between the *a*-axis (i.e., *x*-direction) and the projection of the MA$^+$-ion orientation within the *ab*-plane. Thus, the four possible projection vectors are oriented along $[110], [\bar{1}10], [\bar{1}\bar{1}0]$, and $[1\bar{1}0]$, which respectively correspond to the MA cations lying within the *ab*-plane with the *θ* angles of 45° (for A), 135° (for B), 225° (for C), and 315° (for D). With respect to the *ab*-plane, the MA cations have two symmetric preferred orientations, $\phi = \pm 30°$,[24] where $\phi$ is the tilting angle of the C-N bond axis with respect to the *ab*-in-plane (Figure 1a).

Figure 1b shows the crystal structure of the high-temperature cubic phase composed of the central $PbI_6$-octahedron cage and the surrounding $MA^+$ ions. In the cubic phase which is represented by the $Pm\bar{3}m$ space-group symmetry, the corner-shared $PbI_6$ octahedral frame does not show any tendency of the octahedral tilting along all three directions, *a*, *b*, and *c*. Thus, the cubic phase is represented by $a^0a^0a^0$ in the Glazer's notation. Figure 1c shows the tilted three-dimensional structure of the room-temperature-stable tetragonal phase which belongs to the *I4/mcm* space group. In this tetragonal structure, the $PbI_6$ octahedra do not show any alternative tilting along the *a*- and *b*-axes but exhibit out-of-phase tilting along the *c*-axis, which is in accordance with the $a^0a^0c^-$ tilt pattern in the Glazer's notation.

Let us now consider the difference in the point-group symmetry of the $PbI_6$-octahedron cage between these two relevant phases. In the cubic phase, the $PbI_6$ octahedral network belongs to $O_h$ point group which is characterized by the principal 4-fold rotation axis along the *c*-axis ($C_4$) and the mirror plane perpendicular to this $C_4$ axis ($\sigma_h$; Figure 1b). Owing to the $C_4$ symmetry, a set of the following four distinct orientations of the C-N bond axis is under the same chemical environment: $\{+A, +B, +C, +D\}$. Similarly, a set of the orientations, $\{-A, -B, -C, -D\}$, at a given MA-site is chemically equivalent in the cubic phase. Owing to the $\sigma_h$ symmetry, however, the two orientations having the same $\theta$ angle but with two opposite $\phi$ values (*e.g.*, +A and −A orientations; Figure 1a) are under the same chemical environment. Thus, in the cubic phase, the PbI$_6$-octahedron cage provides all 8 possible orientations of MA, $\{\pm A, \pm B, \pm C, \pm D\}$, with the same chemical environment. This symmetry argument is graphically illustrated in Figure S1 of the Supporting Information.

In the room-temperature-stable tetragonal phase, on the contrary, the PbI$_6$ octahedral network belongs to $D_{2d}$ point group owing to the $a^0 a^0 c^-$ tilt pattern. Thus, the PbI$_6$-cage network is characterized by the S$_4$ improper rotation axis along the *c*-axis (Figure 1c). Because of the S$_4$ improper rotation, a set of the following four distinct orientations of the C-N bond axis (at a given arbitrary MA-site) is under the same chemical environment: $\{+A, -B, +C, -D\}$. Similarly, a set of the orientations, $\{-A, +B, -C, +D\}$, at a given MA-site is chemically equivalent in the tetragonal phase. Consequently, there exist two distinct chemical environments (also, energetically non-degenerate) for the MA$^+$-ion orientation in the tetragonal phase. These two distinct sets of orientations are graphically illustrated in Figure 2: $\{+A, -B, +C, -D\}$ in the upper panel and $\{-A, +B, -C, +D\}$ in the lower panel.

The unit-cell structure of MAPbI$_3$ with the marked four distinct MA-sites is depicted in Figure 3. As displayed in Figure 3a, the four MA dipoles (1,2,1′, and 2′) are located at the same *a-b* plane. When the cell is viewed from the *a*-axis (Figure 3b), the 1$^{st}$ and 2$^{nd}$ MA-sites are on the same *a-b* plane but the 3$^{rd}$ and 4$^{th}$ sites are located at a different *a-b* plane which is (*c*/2) away from the former *a-b* plane along the *c*-axis of the tetragonal *I4/mcm* cell. Thus, the distance between the 1$^{st}$ and 2$^{nd}$ sites (or equivalently, between the 3$^{rd}$ and 4$^{th}$ sites) is given by $R_{12} = \frac{\sqrt{2}}{2}a$, where *a* is the *a*-axis lattice parameter.

**3.2. Two Distinct Modes of Hydrogen-Bonding Interaction.** We have examined the above made proposition on the existence of two non-equivalent chemical environments by examining the MA$^+$-ion orientation in the tetragonal phase on the basis of *ab initio* density-functional theory (DFT) calculations. We used the experimentally obtained lattice parameters $(a = b = 8.80\ \text{Å},\ c = 12.62\ \text{Å})$[35] as the input parameters of our DFT calculations. We have chosen two orientations, +A and −A, to examine the existence of two non-degenerate chemical environments at a particularly chosen MA site. However, our discussion is also valid for other pairs of the MA orientations (*e.g.*, +C & −C). The computed value of *θ* is 45° for both +A and −A orientations [Figure 1a].[24] However, the optimum tilting angle ($\phi$) which corresponds to the minimum in the Kohn-Sham (K-S) energy depends sensitively on the MA$^+$-ion orientation: ~+22° for +A orientation and ~+5° for −A initial orientation. It is interesting to notice that the optimum relaxed tilting angle ($\phi$) for the −A initial orientation is ~+5°, instead of yielding a negative value. This is quite surprising since the input $\phi$ value for the −A orientation (usually $-15^o \leq \phi \leq -10^o$) corresponds to a set of the degenerate orientations, $\{-A, +B, -C, +D\}$ but the relaxed equilibrium $\phi$ value then belongs to a set of the opposite orientations, $\{+A, -B, +C, -D\}$.

This extraordinary result indicates that the particular MA-site that we have chosen in the present DFT calculations strongly prefers the +A orientation to the −A orientation. Let us call this particular site as the 1st MA-site, as shown in Figure 3. Indeed, the calculated K-S energy difference between the two orientations is as large as 45.14 meV per MA-site. Thus, a set of the orientations, $\{-A, +B, -C, +D\}$, does not practically exist at the 1st MA-site though the symmetry argument predicts its existence. Consequently, we end up with a positive equilibrium $\phi$ even if we used a negative input $\phi$ value for the −A orientation. In our calculations of the K-S energy for the +A orientation at the 1st MA-site, we have chosen the site-dependent dipole configuration of [+A,-A,+A,-A] which denotes the MA$^+$-ion orientations of +A, -A, +A, and −A at 1st, 2nd, 3rd, and 4th sites, respectively.  It can be shown that this particular MA$^+$-ion configuration corresponds to the symmetry-allowed lowest energy configuration (See Section 3.4). On the contrary, the [-A,+A,-A,+A] initial configuration was used to evaluate the K-S energy for the −A orientation at the same 1st MA-site. Thus, the K-S energy difference between the two dipole configurations, [+A,-A,+A,-A] and [-A,+A,-A,+A], is 180.56 meV (= 45.14×4).

Let us define the hydrogen-bonding interaction mode that corresponds to the tilting angle ($\phi$) of +22° as the *α*-interaction mode. Similarly, let us denote the hydrogen-bonding interaction mode corresponding to the tilting angle ($\phi$) of +5° as the *β*-interaction mode. Recall that the input $\phi$ value for the *β*-interaction mode is negative although the relaxed value is positive, ~+5°. As mentioned previously, the K-S energy difference between these two tilting-angle states is 45.14 meV per MA-site (i.e., per formula unit). The *α*-interaction mode with $\phi$ value of ~+22° is structurally illustrated in Figure 4a by showing the 1$^{st}$ MA-site (at center) and the surrounding PbI$_6$-octahedron cages. On the other hand, the *β*-interaction mode with $\phi$ value of ~+5° is structurally illustrated in Figure 4b. Herein, the apical (axial) iodine atoms in the PbI$_6$-octahedron cage are denoted by I$_A$, whereas the equatorial iodine atoms are marked with I$_E$. The three hydrogen atoms bonded to the nitrogen (N) atom are denoted by H$_N$ while the three hydrogen atoms connected to the carbon (C) atom are designated by H$_C$.

Among the three H$_N$ atoms that are directly involved in hydrogen bonds, H$_N$(3) atom shows the most prominent difference in the hydrogen-bonding interaction between the *α*- and *β*-modes. In principle, H$_N$(3) is capable of simultaneously interacting with three different equatorial iodine atoms, I$_E$(2), I$_E$(3), and I$_E$(4), in the *α*-interaction mode. On the contrary, H$_N$(3) can interact only with I$_E$(1) in the *β*-interaction mode (See Figure S2 of the Supporting Information). According to the computed bond length and energy (Table 1), three hydrogen bonds are by far outstanding among the 21 possible H-I interactions (11 for the *α*-mode and 10 for the *β*-mode) having a bond critical point where the gradient of the local electron density, $\nabla\rho(\boldsymbol{r})$, is zero. These are: $H_N(1)\cdots I_A(1)$ and $H_N(2)\cdots I_A(2)$ in the *α*-interaction mode and $H_N(3)\cdots I_E(1)$ in the *β*-interaction mode (denoted by dotted red lines in Figure 4). These three hydrogen bonds are named 'the dominant hydrogen bonds.'

We have examined the effect of the hydrogen-bonding mode on the band structure of the tetragonal MAPbI$_3$ cell. The computed band structures are similar to those previously reported by Mosconi *et al.*[23] However, as indicated in Figure 5a, the bandgap at the zone-center Γ-point is significantly affected by the hydrogen-bonding mode. We have further examined the partial density-of-states (PDOS) to resolve the atomic-scale origin of this bonding-mode-dependent bandgap. As indicated in Figure 5b, the conduction-band minimum (CBM) is characterized by the Pb *6p* orbitals, which is irrespective of the hydrogen-bonding interaction mode. On other hand, the valence-band maximum (VBM) is featured by the Pb *6s* and I *5p* orbitals. A detailed analysis of the wavefunction-character indicates that the Pb *6p*–I *5p\** anti-bonding orbital corresponds to the overlapping at the CBM while the Pb *6s*–I *5p\** anti-bonding orbital represents the VBM. It is interesting to notice that in the case of the *α*-interaction mode, the PDOS of the Pb *6p$_z$* orbital at the CBM further penetrates into a lower energy region (down to 1.73 eV above the VBM; Figure 5b). This lowers the CBM value with respect to the VBM, leading to the bandgap reduction in the case of the *α*-interaction mode.

In addition, the Pb *6p$_z$* orbital is expected to show a certain degree of the orbital overlapping with the apical I *5p\** orbital under the *α*-interaction mode. A careful examination of the PDOS indeed shows that the PDOS for the apical I *5p\** (ap) orbital is slightly higher than that for the equatorial I $5p^*$ (eq) orbital near the CBM under the $α$-interaction mode (Figure 5b). Owing to the slightly enhanced Pb $6p_z$–I $5p^*$ orbital overlapping at the CBM, it is predicted that the angle between Pb-(*ap*)I-Pb under the $α$-interaction mode is closer to 180° than the corresponding angle under the $β$-interaction mode. Indeed, our *ab initio* DFT calculations showed that the Pb-(*ap*)I-Pb angle under the $α$-interaction mode ($\omega = 168.6°$) is substantially closer to 180° than the Pb-(*ap*)I-Pb angle under the $β$-interaction mode ($\omega = 160.3°$).

**3.3. Characteristic-Angle-dependent Kohn-Sham Energy.** We have then examined the orientation-dependent K-S energy to assess whether these two hydrogen-bonding modes truly represent a globally stable (minimum) state or not. The orientation of the C-N bond axis is determined by the following three characteristics angles: $\theta$ (azimuthal angle), $\phi$ (tilting angle), and $\chi$ (torsion angle), where $\chi$ defines the rotation angle of the C-N bond axis.[17] For a fixed MA$^+$-ion orientation, both *α*- and *β*-interaction modes have a common azimuthal $\theta$-angle. For instance, $\theta = 45°$ for ±A-orientations (Figure 1a). Thus, we have examined $\phi$- or $\chi$-dependent K-S energy. In Figure 6a, the computed K-S energy is plotted as a function of the tilting angle, $\phi$, which indicates the equilibrium $\phi$-angle for the 1$^{st}$ (or 3$^{rd}$) site is +22° and +5° for *α*- and *β*-interaction modes, respectively. In case of the torsion angle, the K-S energy for the *β*-interaction mode shows its maximum when $\chi$ is at 0° or 120° while the K-S energy shows its minimum when $\chi$ is at 60° (Figure 6b). Contrary to this, the K-S energy shows a reverse trend for the *α*-interaction mode. In this case, a pronounced increase in the K-S energy occurs upon increase in the torsion angle ($\chi$) from 0° to 60° or upon decrease in $\chi$ from 120° to 60° (Figure 6b). This increase in the K-S energy can be understood in terms of the rupture of the relevant hydrogen bonds upon the torsion of the C-N bond axis from its equilibrium $\chi$ values, 0°, 120°, *etc* (Figure 4a).

For each interaction mode, the orientation-dependent energy is described by three characteristic variables, $\theta, \phi$, and $\chi$. Under the thermodynamic equilibrium, the K-S energy should be its global minimum, simultaneously satisfying the two criteria: $(\partial E_{ks}/\partial \theta)_{\phi,\chi} = (\partial E_{ks}/\partial \phi)_{\theta,\chi} = (\partial E_{ks}/\partial \chi)_{\theta,\phi} = 0$ and $(\partial^2 E_{ks}/\partial \theta^2)_{\phi,\chi} = (\partial^2 E_{ks}/\partial \phi^2)_{\theta,\chi} = (\partial^2 E_{ks}/\partial \chi^2)_{\theta,\phi} > 0$ in the three-dimensional $(\theta, \phi, \chi)$–space. Thus, for each interaction mode, the equilibrium $\phi$ and $\chi$ angles deduced from Figure 6 correspond to the globally stable state, not a local minimum.

According to the computed K-S energy shown in Figure 6b, the activation free-energy for the C-N bond rotation is 49.4 meV for the *α*- interaction mode while it is 16.9 meV for the *β*-interaction mode. This suggests that the net hydrogen-bonding strength in the *α*-interaction mode is much stronger than that in the *β*-interaction mode. We will quantitatively examine this important point in Section 3.5. Since the room-temperature thermal energy is 25.7 meV, an effectively free torsional motion of the C-N bond axis is expected in the *β*- interaction mode but not in the *α*-interaction mode. By exploiting the transition-state theory,[36] the frequency of the free torsional rotation is estimated to be: $\nu_\beta = \frac{k_B T}{h} e^{-\Delta G^*_\beta / k_B T} = 3.2 \times 10^{12}$ sec$^{-1} \gg 1$ for the *β*-interaction mode at room temperature. We are in a position to summarize the main difference in the organic MA$^+$-ion orientation between the *α*- and *β*-interaction modes: (i) The equilibrium $\phi$-angle is +22° in the *α*- interaction mode while it is +5° in the *β*-interaction mode. (ii) The orientation relationship of the -NH$_3$ group in the *β*-interaction mode with $\chi = 60°$ (Figure 6b) can be reproduced by the 180° rotation of the N-H$_N$(3) bond axis of the -NH$_3$ group in the *α*-interaction mode ($\chi = 0°$ or 120°) along the *c*-axis. This can be identified by examining the two left-hand side illustrations of Figure 4. On the other hand, both *α*- and *β*-interaction modes have a common $\theta$-angle, as mentioned previously.

**3.4. Remarkably Simplified Dipole Configurations by Considering Structural Symmetry.** Let us begin our discussion by examining conceivable MA-dipole orientations that satisfy the symmetry rule for a given MA-site in the perovskite cell. For this, we have particularly chosen the 1$^{st}$ MA-site among four possible sites in a given perovskite cell (Figure 3). By considering the restriction imposed by the structural symmetry, we have shown that a set of the dipole orientations, $\{+A, -B, +C, -D\}$, is allowed at the 1$^{st}$ MA-site. On the other hand, a set of the orientations, $\{-A, +B, -C, +D\}$, is practically prohibited at the 1$^{st}$ MA-site (Section 3.2). According to the DFT calculations, the +A orientation of the MA$^+$-ion at the 1$^{st}$ MA-site is much more stable than the –A orientation (Section 3.2). One can directly apply this symmetry rule of $\{+A, -B, +C, -D\}$ to the $\pm$C orientation. On the contrary, the reverse is true for $\pm$B and $\pm$D orientations. Specifically, the –B (or –D) orientation is much more stable than the +B (or +D) orientation at the 1$^{st}$ MA-site.

    Let us now extend the above argument to the remaining three MA-sites in the tetragonal unit cell (Figure 3). On the basis of the translational symmetry of the tetragonal MAPbI$_3$ cell, the above symmetry rule can be directly applied to the 3$^{rd}$ site. In other words, the +A (+C) orientation is much more stable than the –A (–C) orientation at the 3$^{rd}$ site, regardless of the hydrogen-bonding interaction mode. Thus, the calculated $\phi$-dependent K-S energy for the 1$^{st}$ MA-site (Figure 6a) can be extended to the 3$^{rd}$ MA-site, as shown in Figure 7a. On the contrary, the reverse is true for the 2$^{nd}$ and 4$^{th}$ sites: the –A (or –C) orientation with a negative tilting angle $\phi$ is much more stable than the +A (or +C) orientation at the 2$^{nd}$ or 4$^{th}$ site, regardless of the interaction mode. The computed $\phi$-dependent K-S energy shown in Figure 7b clearly supports this conclusion. Let us extend this argument of the site-dependent MA orientation to the $\pm$B and $\pm$D orientations. Considering the structural symmetry rule of $\{+A, -B, +C, -D\}$ for the 1$^{st}$ MA-site, one can readily obtain the reverse conclusion for the $\pm$B and $\pm$D orientations. More specifically, the -B (or -D) orientation is much more stable than the +B (or +D) orientation at the 1$^{st}$ and 3$^{rd}$ sites. On the contrary, +B (or +D) orientation is much more stable than the -B (or -D) orientation at the 2$^{nd}$ and 4$^{th}$

sites.

If the MA dipoles are randomly oriented as in the case of the cubic phase, the number of maximum conceivable orientations of the four MA dipoles in the tetragonal unit cell is given by $(2*4)^4$ = 4096 with each orientation represented by characteristic $\theta$ and $\phi$ angles. Herein, '2' takes into account $\pm$ orientations for a fixed $\theta$, '4' represents the four possible values of $\theta$ for a fixed $\phi$, and the power-exponent, 4, takes into account the four distinct MA-sites. Owing to the above symmetry rule of dipole orientations, however, the number of possible orientations of the four MA dipoles in the tetragonal cell can be greatly simplified. To deduce this, assume that the 1st MA-site is occupied by the MA dipole with the +A orientation under the $\alpha$-interaction mode. Then, -A, +B, -C, and +D orientations are allowed at the 2nd MA-site while +A, -B, +C, and -D orientations are allowed at the 3rd MA-site. Likewise, -A, +B, -C, and +D orientations are allowed at the 4th MA-site. Accordingly, we deduce 64 possible dipole configurations if the 1st MA-site is occupied by the MA dipole with the +A orientation. These 64 dipole configurations are listed in Table 2. Similarly, we have 64 distinct dipole configurations for each occupancy of –B or +C or –D dipole at the 1st MA-site. Thus, we have total 256 conceivable dipole configurations in the tetragonal unit cell under the $\alpha$-hydrogen-bonding interaction mode. Exactly the same number of the dipole configurations is allowed for the $\beta$-interaction mode but with a different tilting angle, ~+5°. However, the probability of occupying all four MA-sites by the MA dipoles through the $\beta$-interaction mode is relatively negligible since $\rho_{4\beta} = \{1 + \exp(+4\Delta E_{\alpha\beta}/k_B T)\}^{-1} = 8.85 \times 10^{-4}$, where $\Delta E_{\alpha\beta}$ is equal to 45.14 meV. Considering 4096 maximum possible MA configurations, we have achieved a remarkable simplification in the dipole configurations (256/4096 = 1/16) by considering the structural symmetry of the tetragonal MAPbI$_3$ cell.

**3.5. Evaluation of Individual Hydrogen-Bonding Strength.** We have shown that the tetragonal MAPbI$_3$ perovskite cell under the *α*-interaction mode is much more stable than the same perovskite cell under the *β*-interaction mode with the K-S energy difference of 45.14 meV per MA-site. To quantitatively understand this pronounced mode-dependent structural stability in terms of the strength of the participating hydrogen bonds, we have carefully examined the electron density [$\rho(r)$] and the corresponding Laplacian of charge density [$\nabla^2 \rho(r)$] at all the relevant bond critical points (BCPs) by exploiting the so-called 'quantum theory of atoms in molecules (QTAIM)'.[37] In the QTAIM, the local electronic kinetic-energy density of a given quantum system can be expressed in terms of the first-order density matrix $\rho(r, r')$:

$$G(r) = \frac{1}{2} \nabla \nabla' \rho(r, r')|_{r=r'} \tag{1}$$

where the local kinetic energy term, $G(r)$, is called "the lagrangian kinetic energy density." On the basis of QTAIM, Abramov[38] showed the following expression for $G(r)$ at the BCP, where $\nabla \rho(r_{BCP}) = 0$:

$$G(r_{BCP}) = \frac{3}{10}(3\pi^2)^{2/3}\{\rho(r_{BCP})\}^{5/3} + \frac{1}{6}\nabla^2 \rho(r_{BCP}) \tag{2}$$

Mata et al.[39] further correlated the hydrogen-bonding energy ($E_{HB}$) with $G(r)$ at the BCP using the following relation:

$$E_{HB} = 0.429 \, G(\boldsymbol{r}_{BCP}) \tag{3}$$

The calculated bonding energy and length, together with the associated topological properties [$\rho(\boldsymbol{r}_{BCP})$ and $\nabla^2\rho(\boldsymbol{r}_{BCP})$], are listed in Table 1 for the 10 relevant $H_N \cdots I$ bonds that are directly involved in the hydrogen-bonding interaction.[17] In addition to this, all ten BCPs (five BCPs for each interaction mode) are marked with small circles in Figure S2 of the Supporting Information. The well-known criteria of the hydrogen bonding on the basis of QTAIM[40,41] are (i) $\rho(\boldsymbol{r})$ between 0.002 and 0.034 *a.u.* (atomic unit) and (ii) $\nabla^2\rho(\boldsymbol{r}_{BCP})$ between +0.024 and +0.139 *a.u.* at the BCP, where $\nabla\rho(\boldsymbol{r}_{BCP}) = 0$. A certain degree of the flexibility is effectively allowed in the hydrogen-bonding criteria,[41] especially in the range of $\nabla^2\rho(\boldsymbol{r}_{BCP})$. Thus, all the $H_N \cdots I$ bonds listed in Table 1 practically satisfy the criteria of the hydrogen bonding. Among 10 different hydrogen bonds, three are prominent in their $E_{HB}$ values and bonding lengths (< 2.70Å). Previously, they are named 'the dominant hydrogen bonds' (Section 3.2) and are: $H_N(1) \cdots I_A(1)$ and $H_N(2) \cdots I_A(2)$ in the *α*-interaction mode and $H_N(3) \cdots I_E(1)$ in the *β*-interaction mode.

According to the computed results shown in Table 1, the net difference in the hydrogen-bonding energy [$\Delta E_{HB}$] between the two interaction modes is 43.87 meV (= 381.06−337.19) per formula cell. This clearly supports the previously made conclusion that the tetragonal MAPbI$_3$ perovskite cell under the *α*- interaction mode is much more stable than the same perovskite cell under the $\beta$-interaction mode (Section 3.2). Moreover, the estimated bonding-energy difference by the QTAIM (43.87 meV) nearly coincides with the previously calculated K-S energy difference between the two interaction modes (45.14 meV). As indicated in Eqs. (2) and (3), the computed $E_{HB}$ value by the QTAIM depends on $\rho(\mathbf{r})$ at the BCP. In the DFT, $\rho(\mathbf{r})$ uniquely determines the external potential,[42] thus, the ground-state K-S energy that comprises all the interaction terms including the Hartree energy, the external interaction energy between the nucleus and electrons, and the exchange-correlation energy. Thus, the computed value of $\Delta E_{HB}$ (43.87 meV) by applying the QTAIM can be viewed as the overall K-S energy difference between the two interaction modes (45.14 meV), rather than being interpreted as the difference in the pure hydrogen-bonding interaction energies between the two interaction modes.

## 4. CONCLUSION

On the basis of symmetry argument and DFT calculations, we have made the following conclusions on the tetragonal MAPbI$_3$ perovskite cell: **(i)** There exist two distinct types of the hydrogen-bonding interaction between the MA$^+$-ion and the iodide ions in the PbI$_6$-octahedron network. We named them $\alpha$- and $\beta$-interaction modes. **(ii)** The computed K-S energy difference between these two interaction modes is 45.14 meV per MA-site with the $\alpha$-interaction mode being responsible for the stable hydrogen-bonding network. **(iii)** Based on the individual bonding-energy calculations by exploiting the QTAIM, we have shown that five distinct hydrogen bonds are effective in the tetragonal MAPbI$_3$ under the stable $\alpha$-interaction mode. The net difference in the total hydrogen-bonding energies between these two interaction modes is 43.87 meV per MA-site, which nearly coincides with the K-S energy difference of 45.14 meV. **(iv)** We have further made a remarkable simplification in the MA-dipole configurations by imposing the structural symmetry rule and the tilting-angle-dependent K-S energy to the tetragonal MAPbI$_3$ cell.

## ■ ASSOCIATED CONTENT

● **Supporting Information**

Eight MA-dipole orientations in the cubic MAPbI$_3$ cell; Structural illustration of two distinct modes of the hydrogen-bonding interaction in the tetragonal MAPbI$_3$ cell.

## ■ AUTHOR INFORMATION

**Corresponding Authors**

* **hmjang@postech.ac.kr**

**Notes**

The authors declare no competing financial interest.

## ■ ACKNOWLEDGMENTS

This work was supported by the National Research Foundation (NRF) Grant

funded by the Korea Government (MSIP) (Grant No. 2013R 1A2A2A01068274). Computational resources provided by the KISTI Supercomputing Center (Project no. KSC-2015-C3-016) are gratefully acknowledged.

**\* Figure Captions:**

**Figure 1**. **(a)** Eight preferred orientations of the organic MA$^+$-ion (i.e., C-N bond axis) within the perovskite cavity. Herein, A, B, C, and D represent the projection of the MA$^+$-ion orientations on the *a-b* plane, as measured by the angle $\theta$. The orientation of the MA$^+$-ion with respect to the *a-b* in-plane is represented by the tilting angle $\phi$. According to Quarti *et al.*,[24] the two optimum $\phi$ angles are $\pm 30°$. However, we found that the optimum $\phi$ angle depends sensitively on the hydrogen-bonding interaction mode (See the text for details). **(b)** Crystal structure of the high-temperature cubic $Pm\bar{3}m$ phase viewed from the *c*-axis (left-hand side). The corresponding Pb-I inorganic cage characterized by the C$_4$-rotation axis and the mirror plane $\sigma_h$ perpendicular to the C$_4$ axis (right-hand side). **(c)** Crystal structure of the tetragonal *I4/mcm* phase viewed from the *c*-axis (left-hand side). The corresponding Pb-I inorganic cage characterized by the improper S$_4$-rotation axis (right-hand side).

**Figure 2.** Graphical illustration of the two distinct sets of the MA$^+$-ion orientations (at a given MA-site) in the tetragonal MAPbI$_3$ having *D*$_{2d}$ structural symmetry. **(upper panel)** The central MA$^+$-ion viewed along [110] (upper row) and viewed along [001] (lower row) for a set of the four distinct orientations, $\{+A, -B, +C, -D\}$. **(lower panel)** The central MA$^+$-ion viewed along [110] (upper row) and viewed along [001] (lower row) for a set of the four distinct orientations,

$\{-A, +B, -C, +D\}$.

**Figure 3.** The unit-cell structure of the tetragonal MAPbI$_3$ with the marked four distinct MA-sites. **(a)** The unit-cell structure viewed from the *c*-axis. Herein, the four MA dipoles (1,2,1′, and 2′) lie on the same *a-b* plane. **(b)** The unit-cell structure viewed from the *a*-axis. **(c)** The unit-cell structure viewed from the *b*-axis. **(d)** The structure of tetragonal MAPbI$_3$ unit cell viewed from an arbitrary axis.

**Figure 4.** Illustration of the two distinct modes of the hydrogen-bonding interaction between the MA$^+$-ion and the surrounding PbI$_6$-octahedron cages. **(a)** $\alpha$-interaction mode viewed along [110] (left), viewed along [001] (center), and viewed from an arbitrary axis (right). **(b)** $\beta$-interaction mode viewed along [110] (left), viewed along [001] (center), and viewed from an arbitrary axis (right).

**Figure 5.** The band structure and the partial density of states (PDOS) of the tetragonal MAPbI$_3$ cell for the two distinct modes of the hydrogen-bonding interaction. **(a)** The band structure of the tetragonal MAPbI$_3$ cell under the $\alpha$-interaction mode (left) *versus* the band structure under the $\beta$-interaction mode (right). The *ab initio* band-structure calculations were performed along high-symmetry surface ***k***-vectors of the first Brillouin zone. **(b)** The computed PDOS of the tetragonal MAPbI$_3$ cell under the $\alpha$-interaction mode (left) *versus* the PDOS

under the $\beta$-interaction mode (right). The Pb $6p_z$–I $5p^*$ orbital overlapping at the CBM is reasoned to be closely correlated with the bandgap reduction under the α-interaction mode.

**Figure 6.** The computed Kohn-Sham energy plotted as a function of **(a)** the tilting angle $\phi$ and **(b)** the torsion angle $\chi$, clearly showing the effect of the hydrogen-bonding interaction mode on the two equilibrium angles. Notice that the initially set orientation of the MA$^+$-ion is +A (i.e., positive $\phi$) for the $\alpha$-interaction mode and –A (i.e., negative $\phi$) for the $\beta$-interaction mode.

**Figure 7. (a)** The central MA$^+$-ion with the +A (or +C) orientation (left). The +A (or +C) orientation with a positive equilibrium tilting angle does correspond to the stable MA dipole at the 1$^{st}$ or 3$^{rd}$ MA-site in the tetragonal MAPbI$_3$ cell (right). **(b)** The central MA$^+$-ion with the -A (or -C) orientation (left). The -A (or -C) orientation with a negative equilibrium tilting angle does correspond to the stable MA dipole at the 2$^{nd}$ or 4$^{th}$ MA-site (right). Contrary to the above case, the –B (or –D) orientation represents the stable dipole at the 1$^{st}$ or 3$^{rd}$ MA-site while +B (or +D) orientation corresponds to the stable dipole at the 2$^{nd}$ or 4$^{th}$ MA-site of the tetragonal MAPbI$_3$ cell.

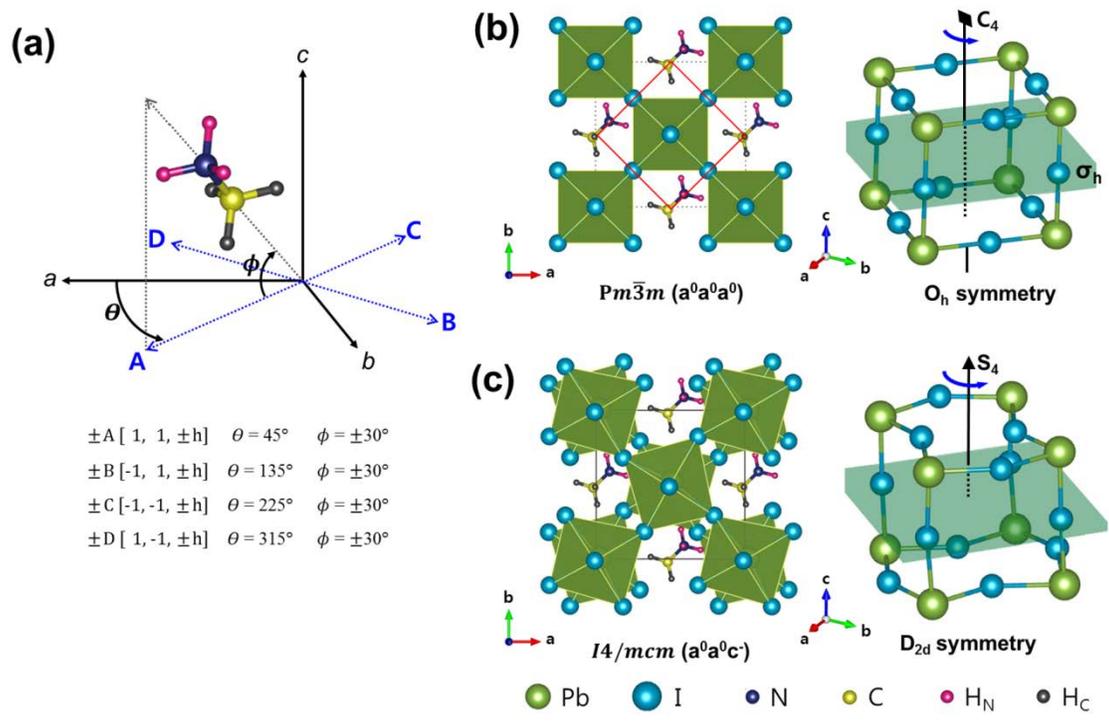

**Figure 1**

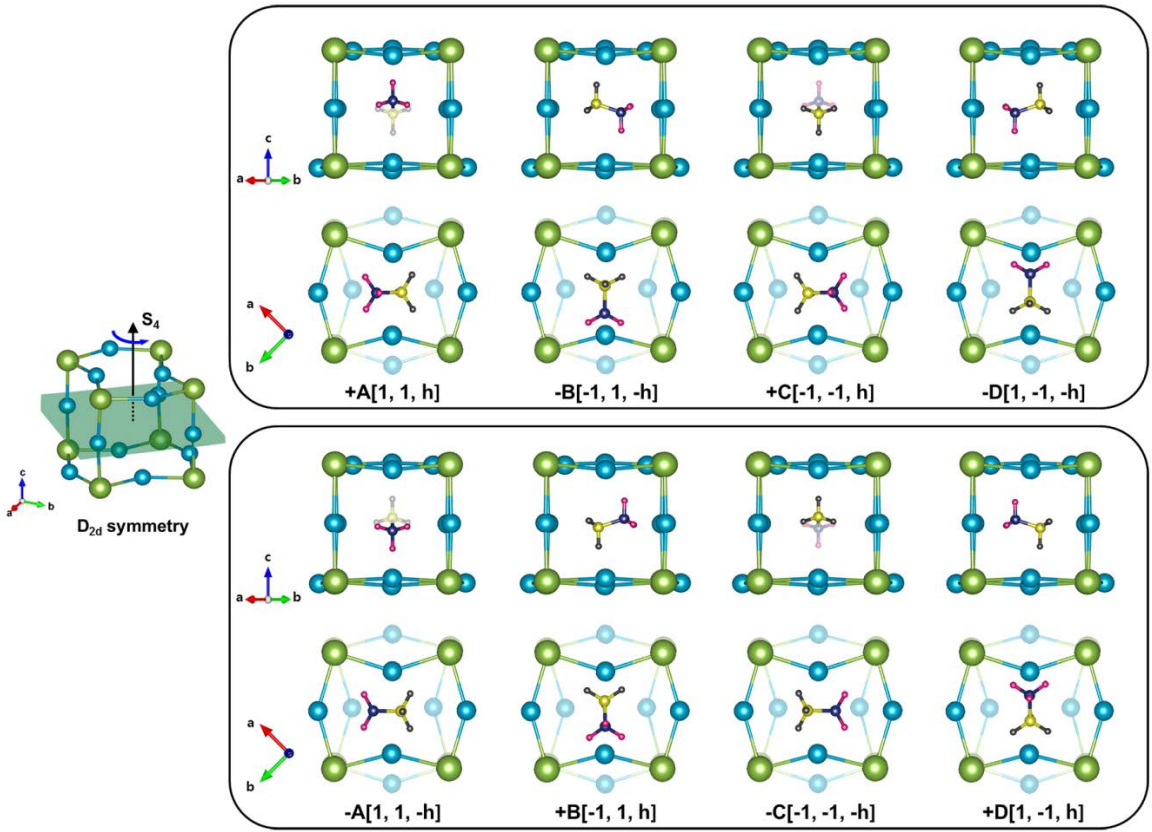

**Figure 2**

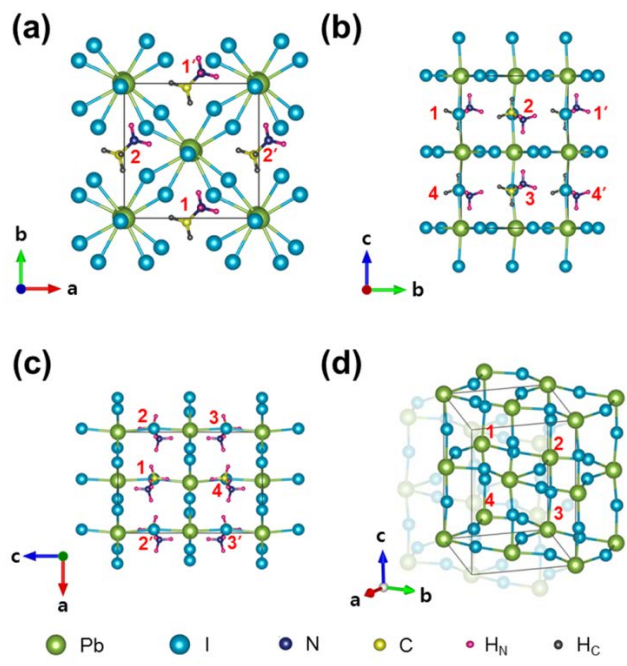

**Figure 3**

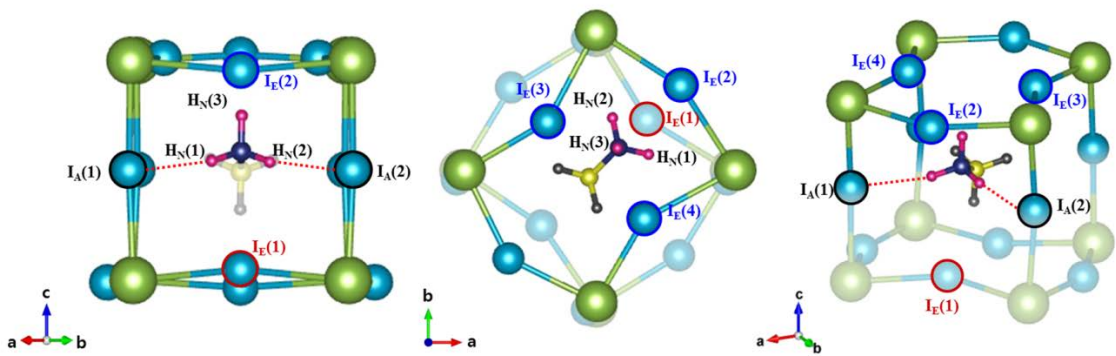

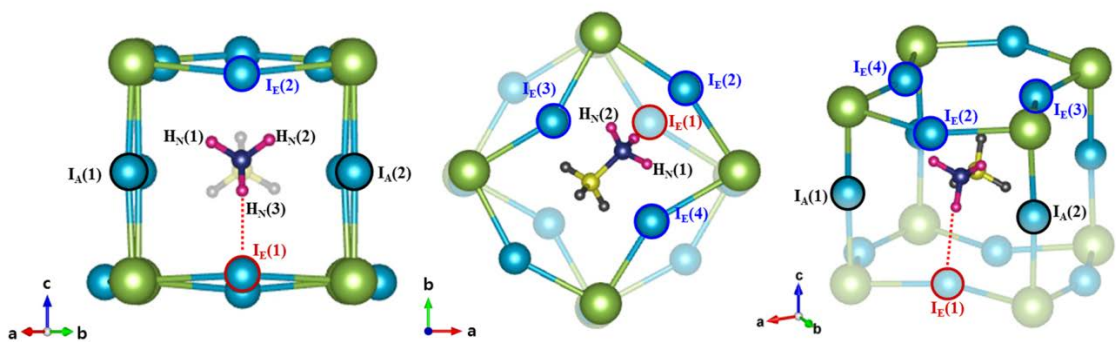

**Figure 4**

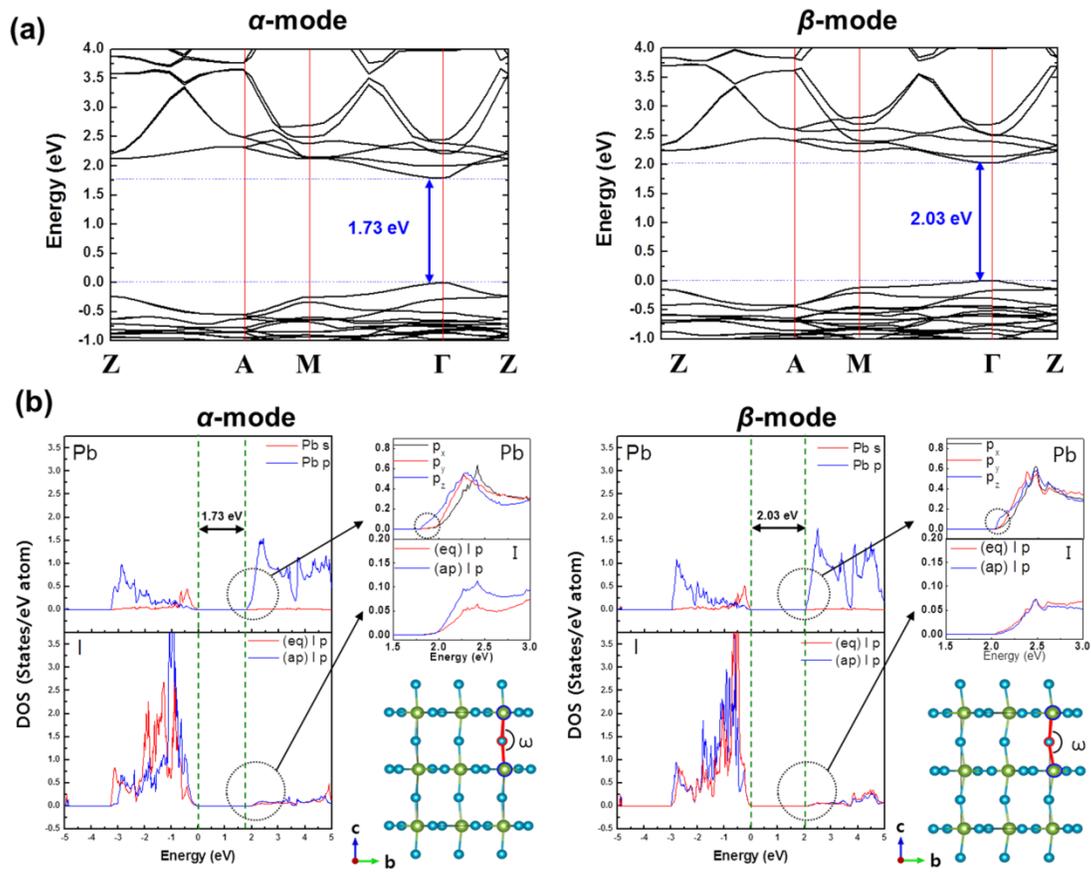

**Figure 5**

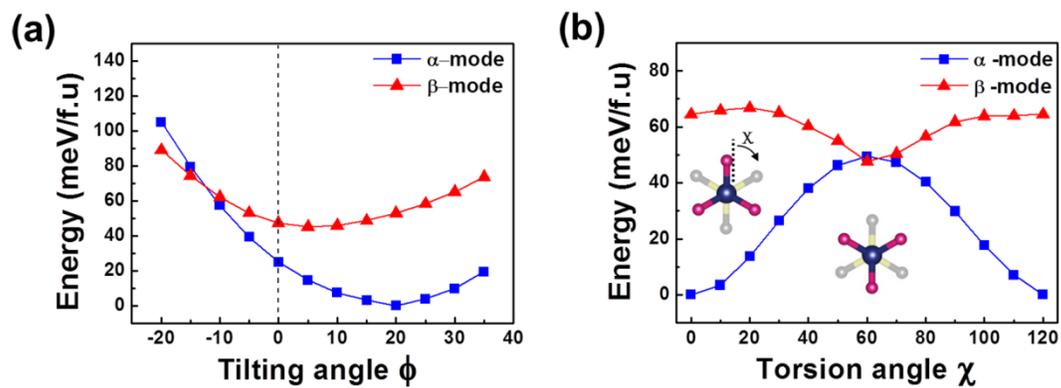

**Figure 6**

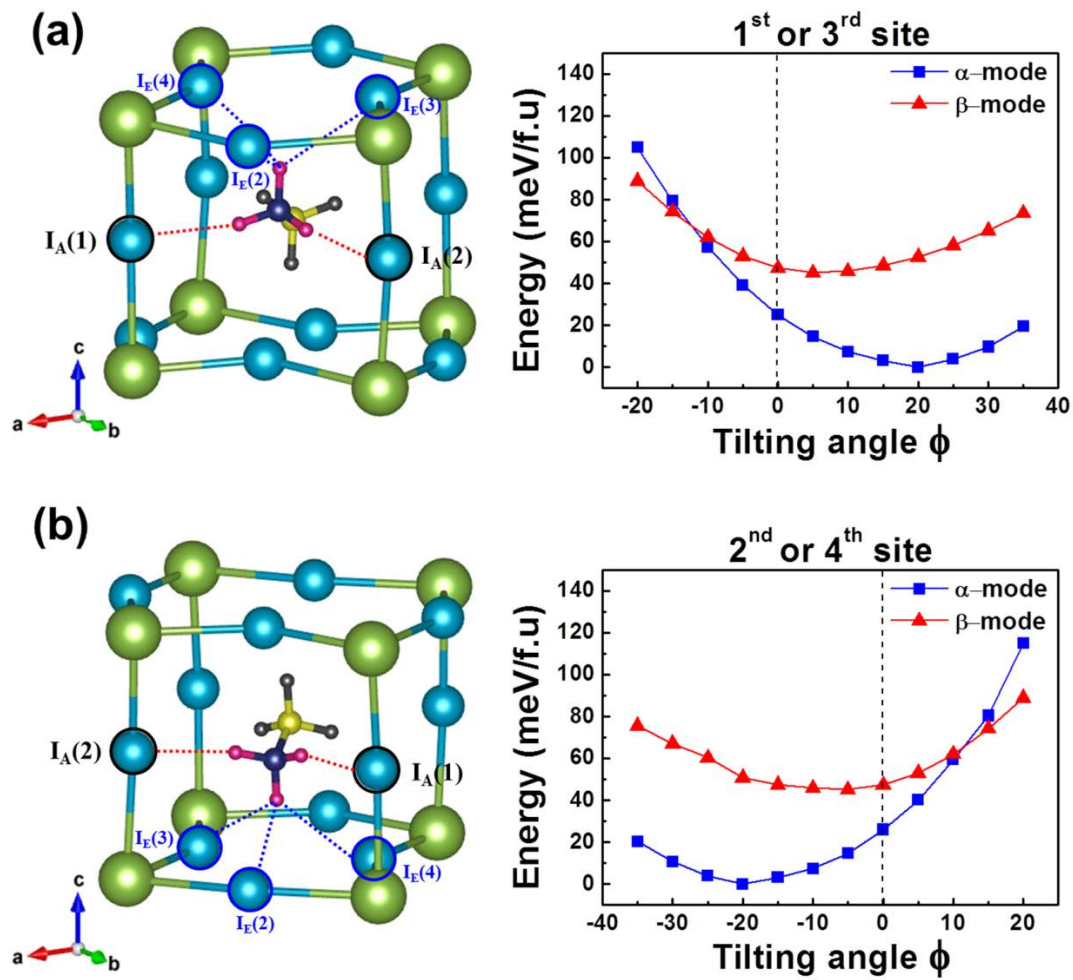

**Figure 7**

**Table 1.** The calculated electronic topological properties, together with the bonding energy, length and angle, for the 10 relevant $H_N \cdots I$ bonds that are directly involved in the two distinct modes of the hydrogen-bonding interaction.

| BCP | $\rho$ (a.u.) | $\nabla^2\rho$ (a.u.) | $E_{HB}$ (meV) | Bonding length (Å) | Bonding angle (°) |
|---|---|---|---|---|---|
| *α*-mode | | | | | |
| $H_N(1) \cdots I_A(1)$ | 0.01747 | 0.03072 | 99.20 | 2.64 | 171.1 |
| $H_N(2) \cdots I_A(2)$ | 0.01706 | 0.03052 | 97.28 | 2.65 | 165.9 |
| $H_N(3) \cdots I_E(2)$ | 0.00882 | 0.02410 | 59.61 | 3.04 | 119.9 |
| $H_N(3) \cdots I_E(3)$ | 0.00994 | 0.02437 | 62.95 | 2.96 | 126.7 |
| $H_N(3) \cdots I_E(4)$ | 0.00958 | 0.02419 | 62.02 | 2.98 | 122.3 |
| Total $E_{HB}$ of *α*-mode | | **381.06**(meV) | | | |
| *β*-mode | | | | | |
| $H_N(1) \cdots I_A(1)$ | 0.00913 | 0.02397 | 59.92 | 2.99 | 132.1 |
| $H_N(1) \cdots I_E(4)$ | 0.01070 | 0.02408 | 64.27 | 2.91 | 133.0 |
| $H_N(2) \cdots I_A(2)$ | 0.01007 | 0.02408 | 62.58 | 2.95 | 133.3 |
| $H_N(2) \cdots I_E(3)$ | 0.01038 | 0.02401 | 63.12 | 2.93 | 132.1 |
| $H_N(3) \cdots I_E(1)$ | 0.01603 | 0.02731 | 87.30 | 2.68 | 175.9 |
| Total $E_{HB}$ of *β*-mode | | **337.19**(meV) | | | |

**Table 2.** 64 symmetry-allowed dipole configurations for the occupation of four distinct MA-dipole sites in the tetragonal MAPbI$_3$ unit cell when the 1$^{st}$ MA-site is occupied by the MA dipole with the +A orientation.

| 1st | 2nd | 3rd | 4th | 1st | 2nd | 3rd | 4th | 1st | 2nd | 3rd | 4th | 1st | 2nd | 3rd | 4th |
|---|---|---|---|---|---|---|---|---|---|---|---|---|---|---|---|
| +A | -A | +A | -A | +A | +B | +A | -A | +A | -C | +A | -A | +A | +D | +A | -A |
| +A | -A | +A | +B | +A | +B | +A | +B | +A | -C | +A | +B | +A | +D | +A | +B |
| +A | -A | +A | -C | +A | +B | +A | -C | +A | -C | +A | -C | +A | +D | +A | -C |
| +A | -A | +A | +D | +A | +B | +A | +D | +A | -C | +A | +D | +A | +D | +A | +D |
| +A | -A | -B | -A | +A | +B | -B | -A | +A | -C | -B | -A | +A | +D | -B | -A |
| +A | -A | -B | +B | +A | +B | -B | +B | +A | -C | -B | +B | +A | +D | -B | +B |
| +A | -A | -B | -C | +A | +B | -B | -C | +A | -C | -B | -C | +A | +D | -B | -C |
| +A | -A | -B | +D | +A | +B | -B | +D | +A | -C | -B | +D | +A | +D | -B | +D |
| +A | -A | +C | -A | +A | +B | +C | -A | +A | -C | +C | -A | +A | +D | +C | -A |
| +A | -A | +C | +B | +A | +B | +C | +B | +A | -C | +C | +B | +A | +D | +C | +B |
| +A | -A | +C | -C | +A | +B | +C | -C | +A | -C | +C | -C | +A | +D | +C | -C |
| +A | -A | +C | +D | +A | +B | +C | +D | +A | -C | +C | +D | +A | +D | +C | +D |
| +A | -A | -D | -A | +A | +B | -D | -A | +A | -C | -D | -A | +A | +D | -D | -A |
| +A | -A | -D | +B | +A | +B | -D | +B | +A | -C | -D | +B | +A | +D | -D | +B |
| +A | -A | -D | -C | +A | +B | -D | -C | +A | -C | -D | -C | +A | +D | -D | -C |
| +A | -A | -D | +D | +A | +B | -D | +D | +A | -C | -D | +D | +A | +D | -D | +D |

TOC graphic

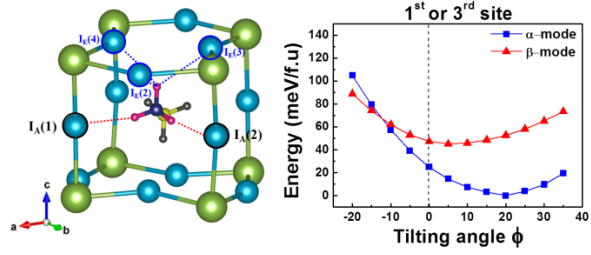

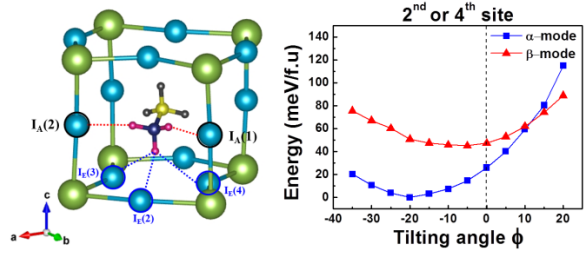

# Supporting Information

## Two Distinct Modes of Hydrogen-Bonding Interaction in the Prototypic Hybrid Halide Perovskite, Tetragonal $CH_3NH_3PbI_3$

**June Ho Lee,[†] Jung-Hoon Lee,[†] Eui-Hyun Kong,[‡] and Hyun M. Jang*[†]**

[†] Department of Materials Science and Engineering, and Division of Advanced Materials Science (AMS), Pohang University of Science and Technology (POSTECH), Pohang 790-784, Republic of Korea.

[‡] Korea Atomic Energy Research Institute (KAERI), Yuseong-Gu, Daejeon 305-353, Republic of Korea.

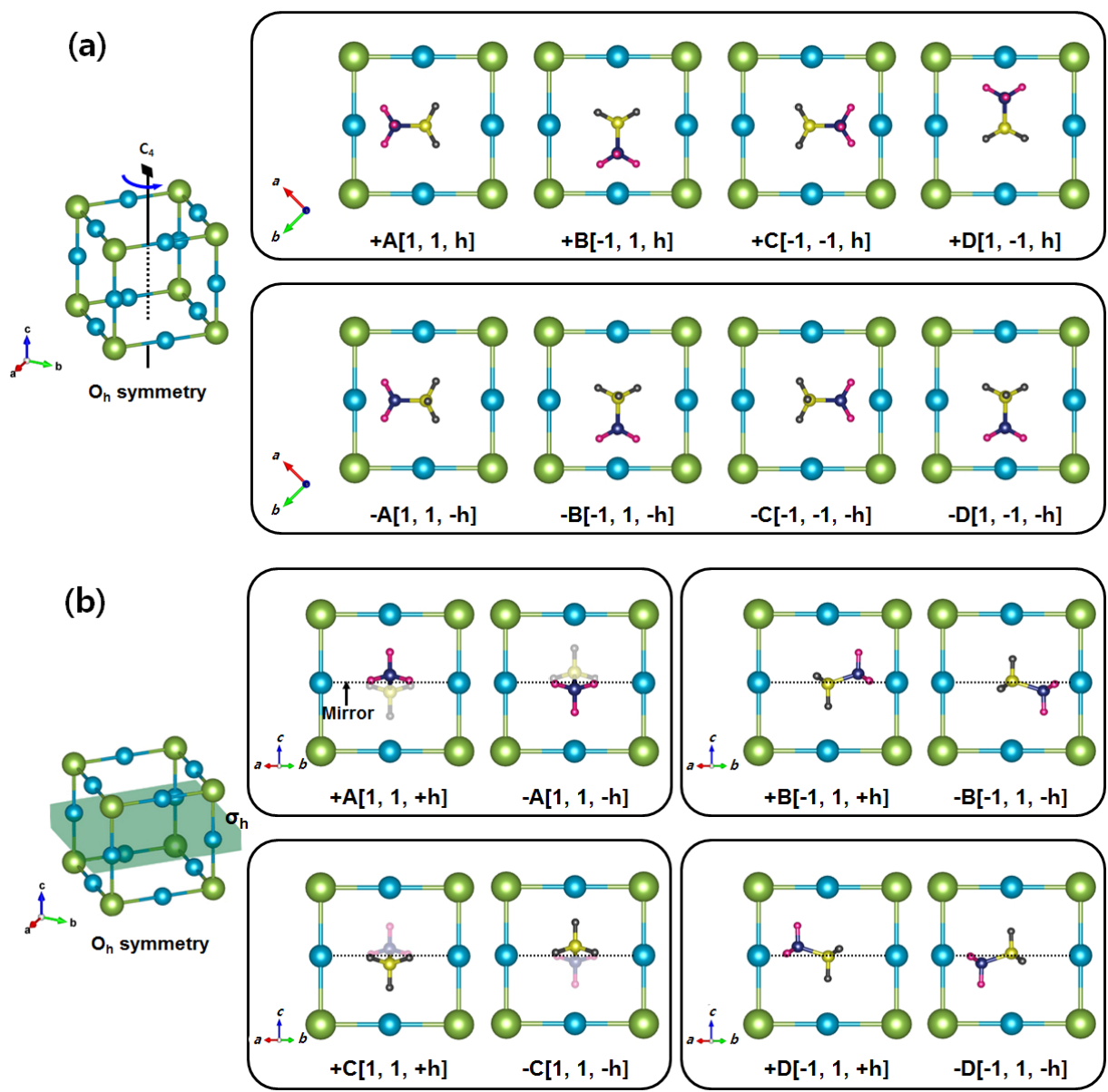

**Figure S1.** Graphical illustration of eight MA-dipole orientations having the same chemical environment in the cubic MAPbI$_3$ cell. **(a)** Two distinct sets of the MA-dipole orientations, $\{+A, +B, +C, +D\}$ and $\{-A, -B, -C, -D\}$, as obtained by the principal $C_4$ rotation operation. **(b)** Four distinct types of the MA-dipole orientations, as obtained by the reflection through the mirror plane ($\sigma_h$) perpendicular to the $C_4$ rotation axis.

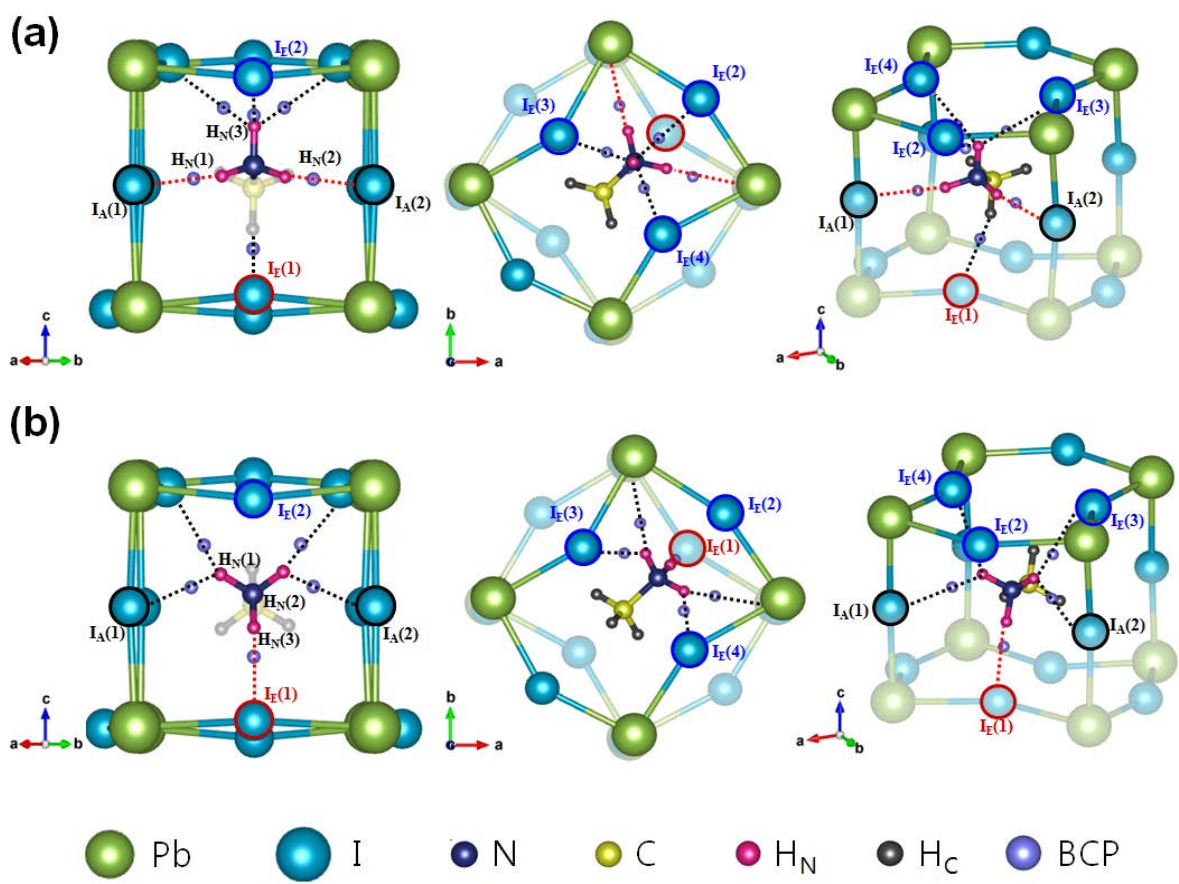

**Figure S2.** Illustration of the two distinct modes of the hydrogen-bonding interaction between the MA$^+$-ion and the surrounding PbI$_6$-octahedron cages. **(a)** $\alpha$-interaction mode viewed along [110] (left), viewed along [001] (center), and viewed from an arbitrary axis (right). **(b)** $\beta$-interaction mode viewed along [110] (left), viewed along [001] (center), and viewed from an arbitrary axis (right). The ten relevant H$_N$⋯I bonds directly involved in the hydrogen-bonding interaction (five per each mode) are denoted by dotted lines. In addition to this, all ten BCPs (five BCPs for each mode) are marked with small circles.